\documentclass[letterpaper, unpublished,11pt]{quantumarticle}
\pdfoutput=1
\usepackage{amsmath}
\usepackage{graphicx}
\usepackage{hyperref}
\usepackage{amsfonts}
\usepackage{amssymb}
\usepackage{braket}
\usepackage{float}
\usepackage{bm}
\usepackage{xcolor}

\begin{document}

\title{The Quantum Alternating Operator Ansatz for Satisfiability Problems}

\author{John Golden}
\affiliation{Information Sciences (CCS-3), Los Alamos National Laboratory, Los Alamos, NM 87545}

\author{Andreas Bärtschi}
\affiliation{Information Sciences (CCS-3), Los Alamos National Laboratory, Los Alamos, NM 87545}

\author{Daniel O'Malley}
\affiliation{Computational Earth Sciences (EES-16), Los Alamos National Laboratory, Los Alamos, NM 87545}

\author{Stephan Eidenbenz}
\affiliation{Information Sciences (CCS-3), Los Alamos National Laboratory, Los Alamos, NM 87545}

\begin{abstract}
We comparatively study, through large-scale numerical simulation, the performance across a large set of Quantum Alternating Operator Ansatz (QAOA) implementations for finding approximate and optimum solutions to unconstrained combinatorial optimization problems. 
Our survey includes over 100 different mixing unitaries, and we combine each mixer with both the standard phase separator unitary representing the objective function and a thresholded version.
Our numerical tests for randomly chosen instances of the unconstrained optimization problems Max 2-SAT and Max 3-SAT reveal that the traditional transverse-field mixer with the standard phase separator performs best for problem sizes of 8 through 14 variables, while the recently introduced Grover mixer with thresholding wins at problems of size 6. 
This result (i) corrects earlier work suggesting that the Grover mixer is a superior mixer based only on results from problems of size 6, thus illustrating the need to push numerical simulation to larger problem sizes to more accurately predict performance; and (ii) it suggests that more complicated mixers and phase separators may not improve QAOA performance.
\end{abstract}

\maketitle
\section{Introduction}\label{sec:intro}
Combinatorial optimization is widely viewed to be a promising application domain for quantum computing, and the Quantum Approximate Optimization Algorithm \cite{Farhi2014} and its generalization the Quantum Alternate Operator Ansatz \cite{hadfield_qaoa} -- both abbreviated as QAOA -- are the most prominent algorithms for this application. 
As theoretical guarantees of QAOA performance are few and far between, we approach QAOA as a heuristic optimization algorithm that may perform very well in practice despite the lack of formal performance guarantees. 
The study of QAOA performance is in its infancy and -- until the arrival of large-scale error corrected quantum computers -- we are limited to numerical experimentation on classical computers. 

QAOA follows a simple iterative approach, evolving an initial state with alternating phase separating and mixing unitaries.
The amount of time (also referred to as the angle) for the phasing and mixing at each iteration (or round) is tuned to increase constructive interference amongst basis states which correspond to high-quality solutions to a target optimization problem, and similarly create destructive interference amongst poor-quality solutions. 
Other studies have shown that QAOA performance depends on the choice of mixing operators, both for unconstrained~\cite{akshay2020reachability} and constrained~\cite{golden2022evidence} optimization; moreover, mixing operators may work better with some phase separators (operators that model the objective function) than others.
Overall, the connection between mixers, phase separators, and QAOA performance is not well understood.
In this paper, we choose Max 2-SAT and Max 3-SAT as our problems of study, which are both unconstrained optimization problems that ask for a truth assignment to variables that maximize the number of clauses in the problem instance; these problems are unconstrained as every possible (binary) assignment of truth values to the values is a feasible solution.
We select a subset of mixing unitaries that spans a wide range of possible higher order mixing terms, with the well-known transverse-field mixer~\cite{Farhi2014} and the recently introduced Grover mixer~\cite{baertschi2020grover} occupying opposite ends of the spectrum.
These mixers are then tested on problems of up to 14 variables.
Previous work~\cite{akshay2020reachability} tested QAOA only with transverse-field and Grover mixers, solving Max 2-SAT and 3-SAT problems with only six variables, and found notable differences in performance with Grover mixer emerging as the winner.
In contrast, our analysis includes a much larger array of mixers and phase separators, and tests $k$-SAT problems up to size $n=14$.

Our main finding is that mixer performance depends on problem size, and that the transverse-field mixer is a poor choice for small problems, but outperforms all other mixers on problems with $\ge$ 10 variables.
We also find that relative performance depends on the exact choice of metric for performance. 
In particular, we compare approximation ratio vs. ground state probability, and find that certain QAOA variations can have a relatively high approximation ratio while having low ground state probability.

\section{QAOA Review}
Using QAOA to solve an optimization problem, defined here by a cost function $C(x)$ on binary strings, requires several choices: an initial state $\ket{\psi_0}$, a mixer Hamiltonian $H_M$, a phase separator Hamiltonian $H_P$, and a set of $2p$ parameters (commonly referred to as angles) $\bm{\beta} = \{\beta_1,\ldots,\beta_p\}$ and $\bm{\gamma} = \{\gamma_1,\ldots,\gamma_p\}$.
One then uses a quantum computer to apply the mixing and phase separating unitaries over $p$ rounds to prepare the state
\begin{equation}
    \ket{\psi_p} = e^{-i\gamma_p H_P}e^{-i\beta_p H_M}\cdots e^{-i\gamma_1 H_P}e^{-i\beta_1 H_M}\ket{\psi_0}.
\end{equation}
The goal is to choose the above parameters so that sampling from $\ket{\psi_p}$ is likely to return a state which is a good (or ideally optimal) solution to the optimization problem. 
This is generally accomplished by fixing $\ket{\psi_0}, H_M, H_P$ and using classical optimization techniques to modify $\bm{\beta}$ and $\bm{\gamma}$ in order to maximize (or minimize) $\braket{\psi_p|H_C|\psi_p}$, where $H_C$, or cost Hamiltonian, encodes the optimization problem at hand, $H_C\ket{x} = C(x)\ket{x}$.

In this work we study unconstrained problems involving $n$ binary variables, and we use the standard initial state $\ket{\psi_0}=\ket{+}^{\otimes n}$.
Recent work has argued in favor of using classical algorithms to generate initial states which are weighted in favor of likely good solutions~\cite{egger2021warm, tate2020bridging}, and the effect of this ``warm-start'' approach in conjunction with different mixers and phase separators, particularly for small number of rounds, is worth future study.

Mixers for unconstrained optimization can take many forms. 
Two well-known examples are the transverse-field mixer, $\sum_{i=1}^n X_i$, originally introduced in~\cite{Farhi2014}, and the Grover mixer, $\ket{\psi_0}\bra{\psi_0}$, originally introduced in the context of $k$-SAT in~\cite{akshay2020reachability} and generalized in~\cite{baertschi2020grover}.
For unconstrained problems, where all $n$-bit strings represent feasible solutions, the space of possible mixer Hamiltonians is very large.
In this work we restrict ourselves to mixers composed of products of Pauli $X$ operators which are symmetric across qubits and have constant coefficient across terms.
This choice encompasses many of the mixers previously studied in the literature, e.g. both the transverse-field and Grover mixers, however the vast majority of this subspace remains unexplored.
This subspace of mixers can be parameterized by sets of integers, where each integer indicates the different degrees of $X$ products to include in the sum.
For example, the set $\{1\}$ is equivalent to the transverse-field mixer defined above, and the set $\{1,2\}$ gives $\sum_{i=1}^n X_i+\sum_{i<j}^n X_i X_j$, and so on.
In general, a set $W$ of $k$ distinct integers $\{w_1,\ldots,w_k\}$, $1\le w_i \le n$, defines the mixer Hamiltonian $H_M(W)$ by
\begin{equation}
    H_M(W) = \sum_{j=1}^{k} \sum_{i_1<i_2<\ldots<i_{w_j}} X_{i_1}X_{i_2}\cdots X_{i_{w_j}}.
\end{equation}
If $W = \{1,2,\ldots,n\}$, then $H_M$ is equivalent to the Grover mixer (modulo constant factors) when applied to the initial state $\ket{+}^{\otimes n}$.

We study the traditional objective-value phase separator~\cite{Farhi2014}, $H_P = H_C$, as well as the threshold-based phase separator~\cite{golden2021thresholdbased}, $H_P = \theta(H_C-t)$, where $t$ is a user-defined threshold and $\theta(x)$ is the Heaviside step function.
Throughout the rest of this work, we will refer to a specific combination of mixer and phase separator as a QAOA \emph{implementation}, and we will often abbreviate these by appending the name of the mixer with either -Th or -Obj to indicate threshold or objective-value phase separators, respectively.

Many different techniques exist for determining the angles $\bm{\beta}$ and $\bm{\gamma}$, including fixed angles~\cite{wurtz2021fixed}, interpolation~\cite{cook2020kVC, Wurtz_2022}, and unsupervised learning~\cite{Moussa_2022}.
In this work we follow an angle-finding strategy which combines basin-hopping starting at an initial point extrapolated from optimal angles at $p-1$ rounds.
The threshold-based phase separator also requires tuning the parameter $t$, however in practice this is not computationally onerous, and in all cases where we report -Th results they reflect the optimal choice of $t$ for the given number of rounds.
The specifics of the angle- and threshold-finding approaches are detailed in~\cite{golden2022evidence}.
This particular strategy places an emphasis on finding very good angles rather than reducing the computational cost of angle-finding.
All of the mixers and phase separators we include here can be implemented in polynomial depth~\cite{golden2022evidence}.
Therefore, we use the number of QAOA rounds as a proxy for computational complexity when comparing relative performance of different implementations.

Throughout this work we will employ three different metrics for performance.
The first metric is the number of QAOA rounds $p$ necessary to have $\ket{\psi_p}$ be composed of only optimal solutions.
This metric is important in the context of~\cite{akshay2020reachability}, as their ``reachability deficits'' are defined in relation to this goal.
The second metric is the approximation ratio, which is defined as $\braket{H_C}/\text{max }C(x)$.
This metric is particularly useful when viewing QAOA as an approximate optimization scheme, and is commonly used in classical optimization analysis.
The third metric is ground state probability (GSP), the probability of returning a ground state, i.e. optimal solution to $C(x)$, when observing $\ket{\psi_p}$.
GSP is useful as a bridge between the first two metrics, i.e. in scenarios where there is a strong emphasis placed on optimal (as opposed to nearly-optimal) solutions, but still in an approximate context (e.g. low number of rounds).
We will show in Sec.~\ref{sec:results} that different QAOA implementations can perform differently -- sometimes dramatically so -- based on which of these metrics one is interested in.

\section{Results}\label{sec:results}
\subsection{Hard \texorpdfstring{$k$}{k}-SAT instances for \texorpdfstring{$n>6$}{n>6}}

\begin{figure*}
    \centering
    \includegraphics[width=\textwidth]{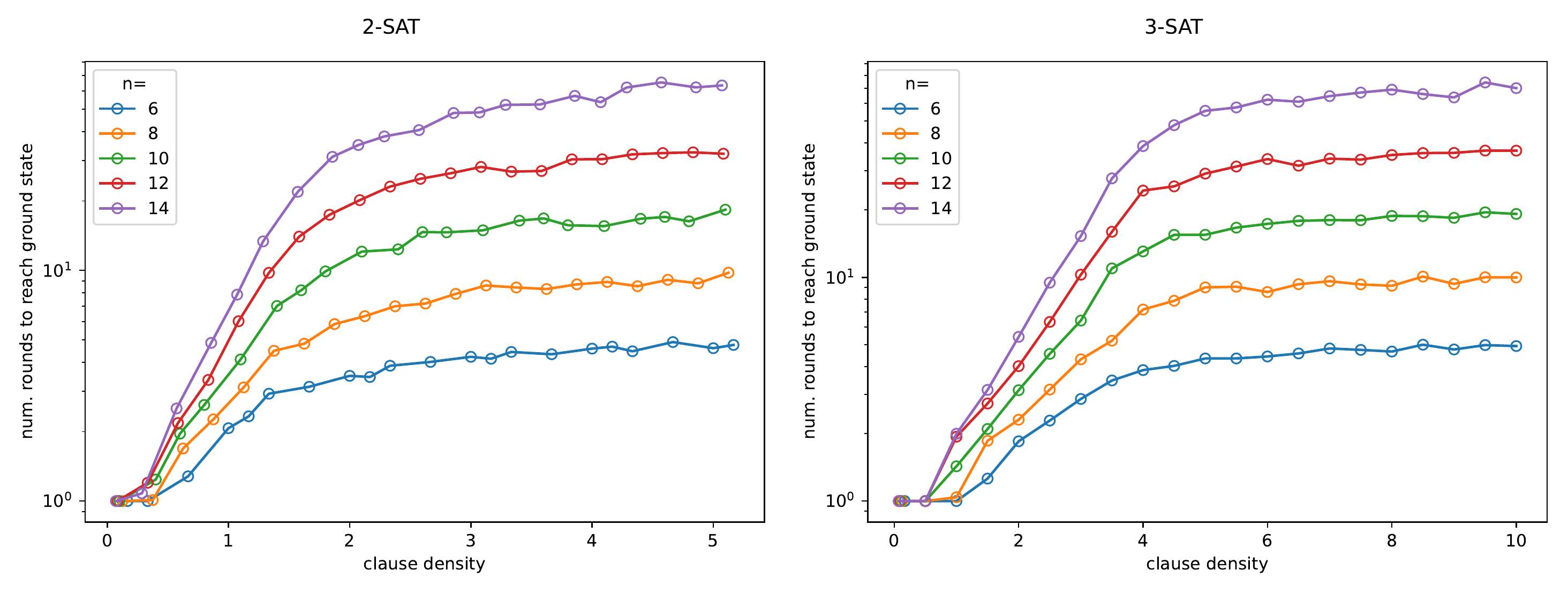}
    \caption{Number of rounds with Grover-Th necessary to achieve an approximation ratio of 1 as a function of clause density. Data represents mean results over 100 random $k$-SAT instances. Setting the clause density at 3,6 ensures ``hard'' problems for 2-SAT, 3-SAT (respectively).}
    \label{fig:hardness}
\end{figure*}

We study QAOA performance in solving conjunctive normal form satisfiability optimization problems.
Such problems are generally phrased in terms of $n$ Boolean variables organized into $m$ clauses of length $k$, and are referred to as $k$-SAT.
In our context we are specifically interested on the Max $k$-SAT variation, that is, the goal is to find the variable assignment(s) which satisfy the maximal number of clauses (and will simply be referred to as $k$-SAT from here on).
The computational resources necessary to solve this problem are known to depend on the clause density, $\alpha=m/n$, for both classical~\cite{zhang2001phase} and quantum~\cite{akshay2020reachability} solvers.
In general, finding the maximum satisfiability for a random $k$-SAT instance with low clause density requires less computational resources than one with a high clause density.
Furthermore, there exists a phase transition, where the difficulty ramps up sharply.
In the case of QAOA case, difficulty here is measured as the number of QAOA rounds necessary to reach a state composed only of optimal solutions.
This phase transition was shown in~\cite{akshay2020reachability}, but only at $n=6$.

Here we extend this analysis to higher $n$.
In Fig.~\ref{fig:hardness} we show how the difficulty scales with clause density for both 2- and 3-SAT.
These results show that the clause densities associated with the hard regime gradually increase.
However, the rate of increase appears quite small, and for problems up to $n=14$ it is sufficient to set the clause density at 3 for 2-SAT and 6 for 3-SAT to be in the hard regime.

\subsection{Broad mixer survey}

\begin{figure*}[t]
    \centering
    \includegraphics[width=\textwidth]{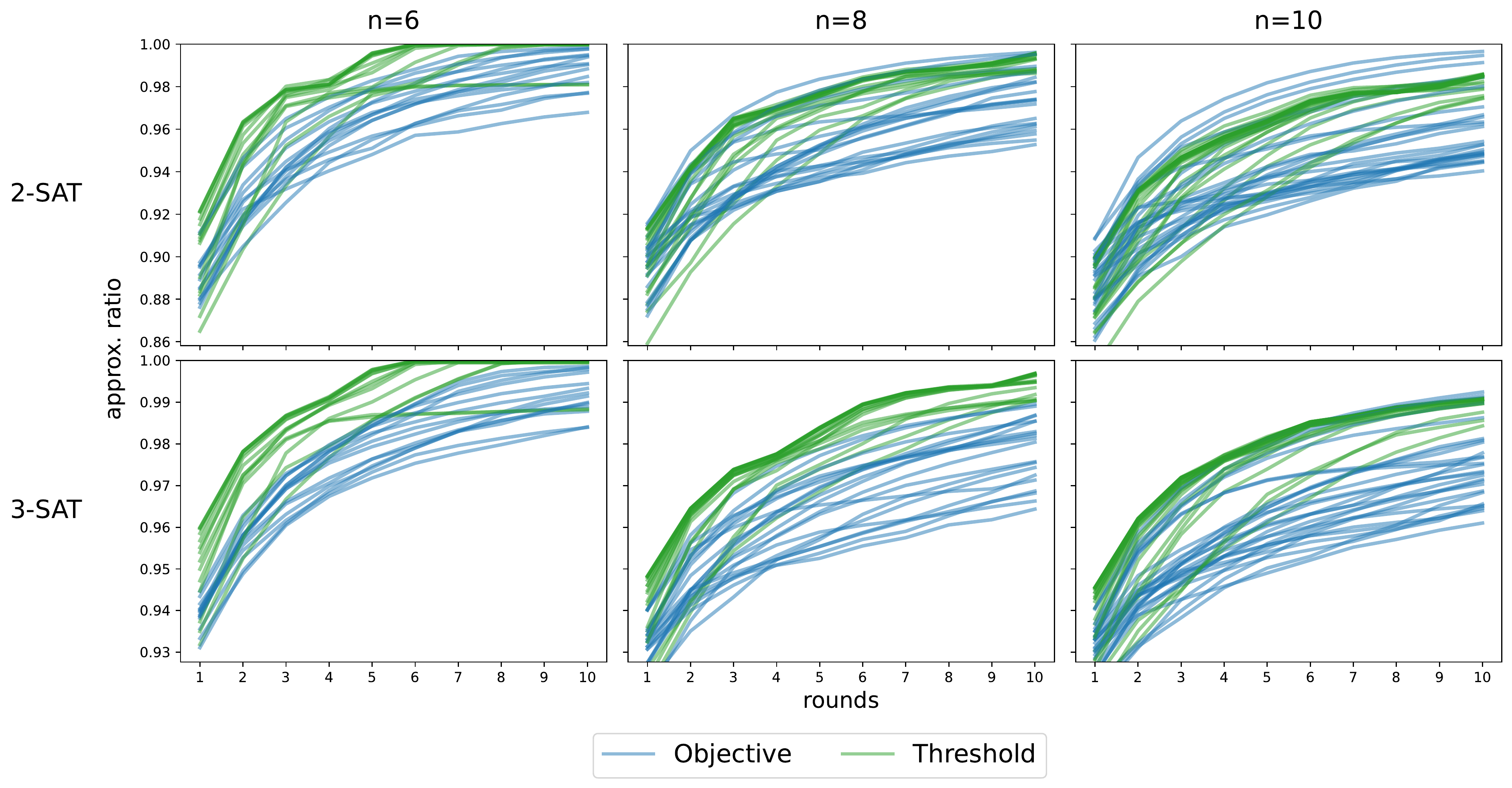}
    \caption{Approximation ratio as a function of rounds for the $6(n-1)$ mixers described in Eq.~\ref{eq:mixer-choice}, with both -Obj (blue) and -Th (green) phase separators. Data averaged over 10 random problem instances per $n$. For $n=6$, the threshold QAOA implementations tend to perform better than those with objective value-based phased separators. However, certain -Obj implementations improve steadily with $n$, and the highest performing implementations at $n=10$ are -Obj for both 2- and 3-SAT.}
    \label{fig:low-n}
\end{figure*}

Having now ensured we are studying difficult random 2- and 3-SAT instances, we can now compare the performance of a wide variety of QAOA implementations at different $n$.
This will answer two questions: which QAOA implementation performs the best at a given $n$, and does relative performance depend on $n$?
For this section we will focus solely on the performance metric of approximation ratio as a function of number of QAOA rounds $p$, covering up to $p=10$.

The subspace of mixers as described in Sec.~\ref{sec:intro} contains $2^n-1$ different mixers (ignoring $W=\{\}$), so testing the performance of every mixer is computationally infeasible for $n>6$.
Instead, we focused on the following collection of $3(n-1)$ mixers:
\begin{equation}\label{eq:mixer-choice}
W \in \begin{cases}
    \{1,\ldots,i\} \text{ for } i=2,\ldots,n,\\
    \{i,\ldots,n\} \text{ for } i=2,\ldots,n-1,\\
    \{i\} \text{ for } i=1,\ldots,n.
    \end{cases}
\end{equation}
These mixers capture a wide range of different mixing approaches.
For example, the transverse-field mixer $W=\{1\}$ mixes solution vectors which are Hamming distance 1 apart, while the mixer $W=\{n\}$ mixes a solution with its inverse (Hamming distance $n$), and the Grover mixer $W=\{1,\ldots,n\}$ mixes all states equally.
The choice of mixers in Eq.~\ref{eq:mixer-choice} interpolates smoothly between these different extremes.
Furthermore, for each mixer, we test both the -Obj and -Th phase separators.
Thus for each $n$ we study $6(n-1)$ different QAOA implementations.
We tested these implementations via highly optimized QAOA statevector simulation code on a cluster of 12 NVIDIA RTX A6000s, each with 48GB RAM.
A single $n=10$ problem instance took $\mathcal{O}(1)$ day to simulate 10 rounds of the 54 different implementations.  

Fig.~\ref{fig:low-n} shows the relative performance of these QAOA implementations for $n=6,8,10$, as measured by approximation ratio for $p\le10$.
Due to the large number of implementations, it is difficult to precisely yet succinctly summarize the differences in performance for all of the mixers, however some broad trends do emerge.
Most apparent from visual inspection is that QAOA implementations with the threshold-based phase separator (i.e. green lines) tend to achieve higher approximation ratios than those with the objective-based phase separator (i.e. blue lines) for $n\le 10$.
The best performing -Th implementations all tended to perform very similarly, this can be seen in the thick green lines (which represent many overlapping -Th results).
These high-performing -Th implementations generally involved a large number of $X$'s, including Grover-Th as well as e.g. $\{n/2,\ldots,n\}$ and $\{1,\ldots,n/2\}$.
``Individual'' mixers, e.g. $\{1\}$-Th, performed the worst amongst the -Th implementations.
The inverse is true for the -Obj implementations, that is, ``small'' mixers such as $\{1\}$ and $\{1,2\}$ tended to perform the best, while mixers that included many products (notably including Grover-Obj) performed the worst.

\subsection{Transverse vs. Grover up to \texorpdfstring{$n=14$}{n=14}}

\begin{figure*}[t]
    \centering
    \includegraphics[width=\textwidth]{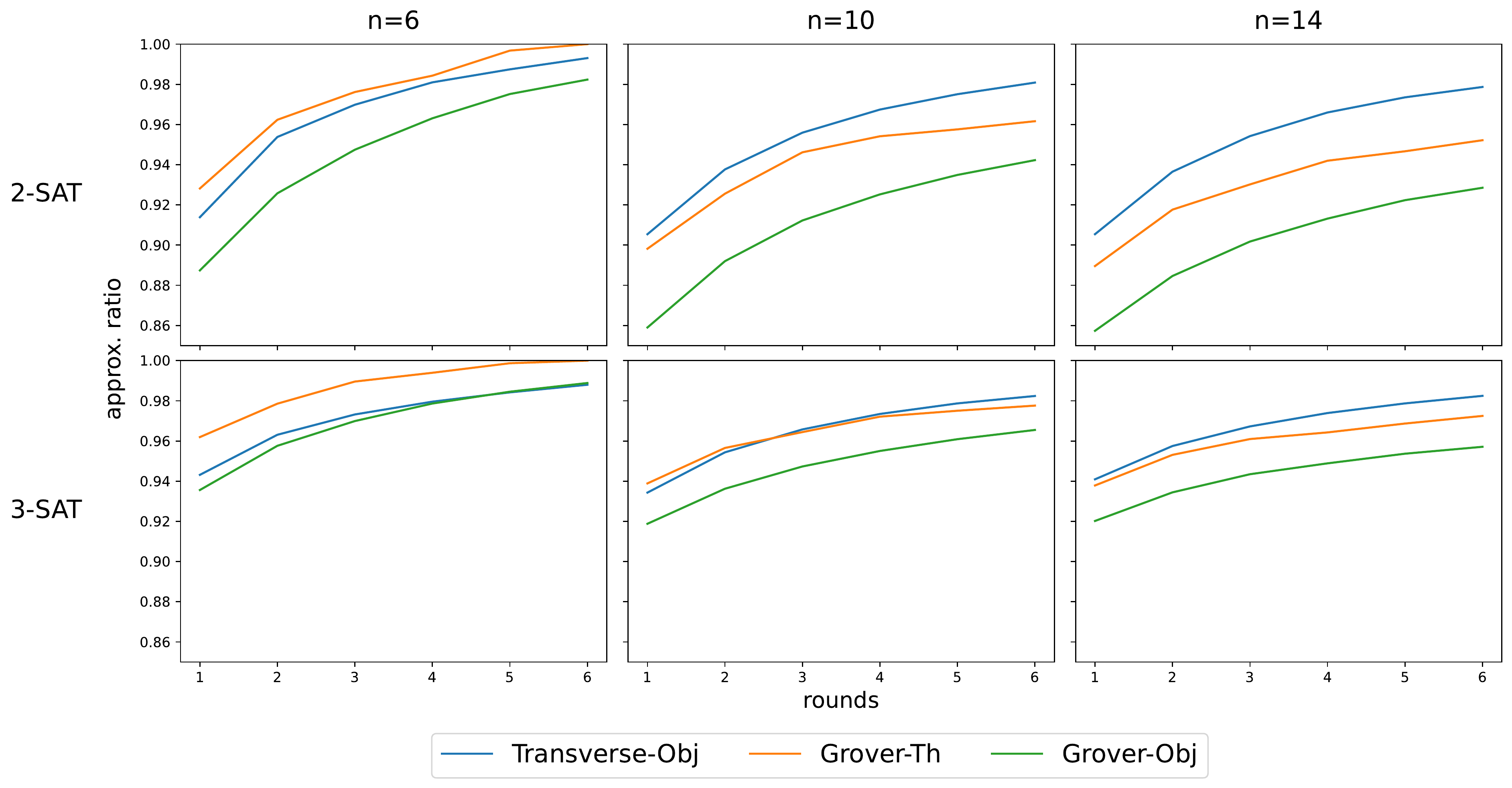}
    \caption{Ground state probability as a function of rounds for Transverse-Obj, Grover-Th, and Grover-Obj for rounds $\le 6$ across a range of $n$ for 2- and 3-SAT. Data averaged over 10 random problem instances. Transverse-Obj significantly outperforms the others for $n\ge 10$.}
    \label{fig:n6-10-14-ar}
\end{figure*}

The most surprising result of Fig.~\ref{fig:low-n}, however, is that as $n$ increases a small number of objective-based separators steadily improve, and by $n=10$ the best performing implementations are $\{1\}$-Obj and $\{1,2\}$-Obj (for both 2- and 3-SAT).
Threshold-based implementations still tend to perform better.
For example, in the $n=10$ 3-SAT case, 23 out of the top 26 implementations are threshold-based (for 2-SAT the proportion is 22/26).
However, the $\{1\}$-Obj and $\{1,2\}$-Obj implementations (as well as the $\{2\}$-Obj implementation in the case of 2-SAT) are clearly achieving higher approximation ratios.
In Fig.~\ref{fig:n6-10-14-ar} we specifically compare the relative performance of $\{1\}$-Obj, i.e. the transverse-field mixer, against both Grover-Th and Grover-Obj up to $n=14$.
Grover-Th is included in the comparison as it is representative of the best-case -Th performance, and Grover-Obj is included as it is the primary mixer of interest in~\cite{akshay2020reachability}.
Here we see even more clearly how the superior performance of Transverse-Obj is only apparent as $n$ increases.
This result stands in contrast to the conclusions drawn in~\cite{akshay2020reachability}, where they observed that Grover-Obj outperformed Transverse-Obj at $n=6$ and implied that this result held for more general $n$.

\subsection{Approximation Ratio vs. Ground State Probability}

\begin{figure*}[t]
    \centering
    \includegraphics[width=\textwidth]{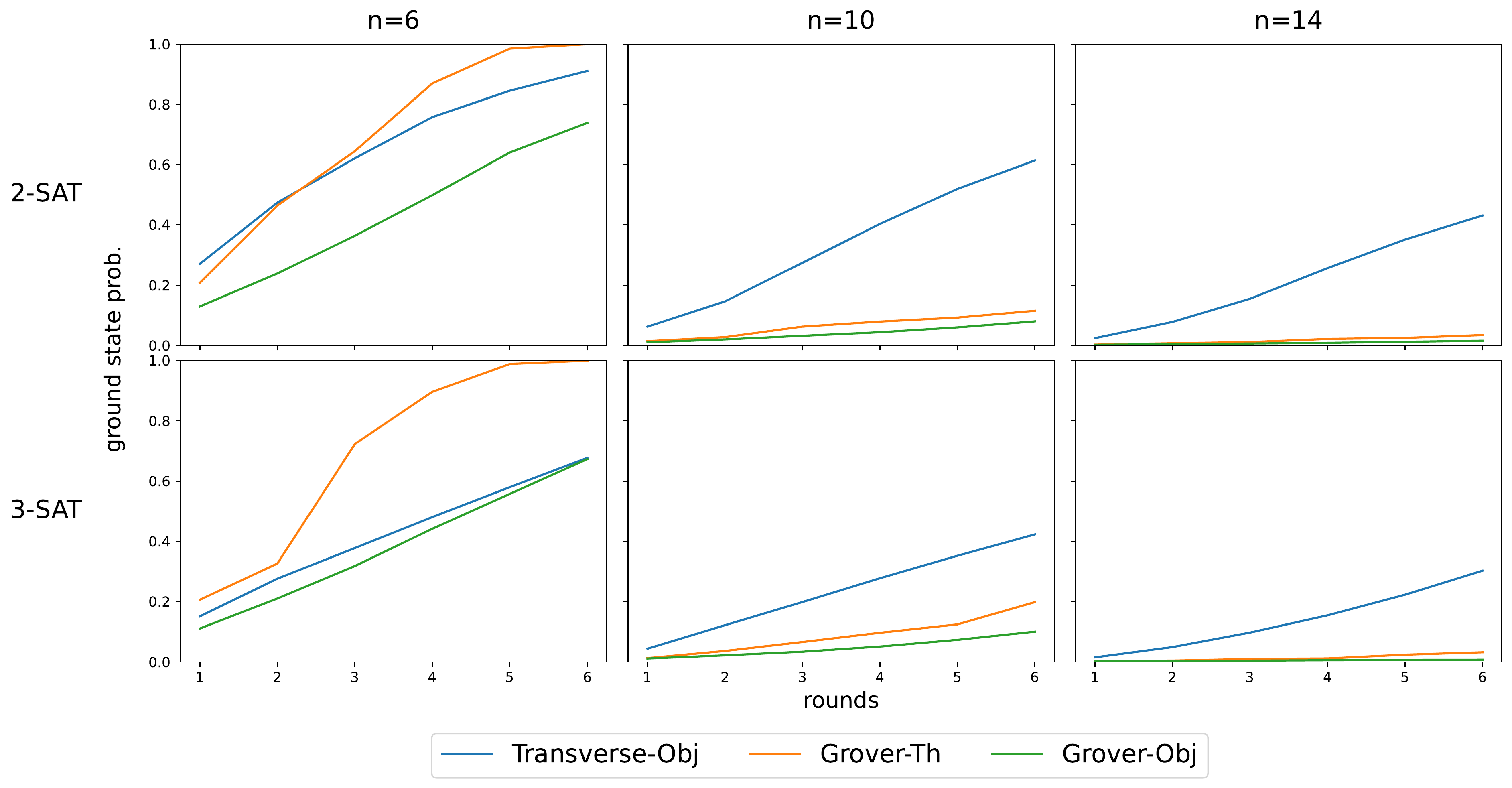}
    \caption{Ground state probability as a function of rounds for Transverse-Obj, Grover-Th, and Grover-Obj for rounds $\le 6$ across a range of $n$ for 2- and 3-SAT. Data taken from the same 10 random instances as Fig.~\ref{fig:n6-10-14-ar}. Transverse-Obj significantly outperforms the others for $n\ge 10$.}
    \label{fig:n6-10-14-gsp}
\end{figure*}

We now study the relative performance of Transverse-Obj, Grover-Th, and Grover-Obj when measured by ground state probability (GSP), see Fig.~\ref{fig:n6-10-14-gsp}.
This particular choice of regime ($p\le 6$ up to $n=14$) and performance metric is meant to highlight the practical differences between QAOA implementations in an approximate optimization context.
It is important to emphasize here that the angles $\bm{\beta}, \bm{\gamma}$ are still chosen to optimize $\braket{H_C}$, as opposed to GSP, since GSP is only measurable once one has already solved the optimization problem of interest. 
So it is quite striking that the different QAOA implementations isolate the optimal states to vastly differing degrees.
At $n=6$, Grover-Th quickly creates a $\ket{\psi_p}$ dominated by optimal states.
For these small problems, setting the threshold sufficiently high to only filter out the optimal states is the best strategy for both improving $\braket{H_C}$ as well as GSP.
In this case, Grover-Th is equivalent to a direct Grover search for optimal states.
Meanwhile, Transverse-Obj and Grover-Obj both improve steadily with each round.
However, by $n=10$, both of the Grover implementations feature drastically lower GSPs, while Transverse-Obj has only decreased slightly, and the story continues at $n=14$.
Comparing the results of Figs.~\ref{fig:n6-10-14-ar} and \ref{fig:n6-10-14-gsp}, which are obtained from the same problem instances and $\bm{\beta}$, $\bm{\gamma}$, we can infer that for $n\ge10$ the Grover mixer, with both -Obj and -Th phase separators, improves the approximation ratio by increasing the amplitude of a large number of states with sub-optimal objective values.
Meanwhile, the Transverse-Obj implementation is much more effective at increasing the amplitude of optimal states.

\section{Conclusions and Future Directions}\label{sec:conclusion}

In this work we have conducted the first large-scale comparison of QAOA mixers and phase separators for unconstrained optimization problems.
The central conclusion of this work is that the traditional transverse-field mixer with objective value phase separator is the most performant QAOA implementation as problem size grows.
This stands in contrast to prior work~\cite{akshay2020reachability}, which conjectured that mixing Hamiltonians that mixed across a wide range of states were more effective.
Instead, we argue the opposite: the effectiveness of the transverse-field mixer is due to the fact that it predominantly mixes states which are close in Hamming distance.
This is useful because states which are nearby in Hamming space often have similar objective values for a random $k$-SAT instance.
Used in conjunction with the objective value phase separator, Transverse-Obj is thus more able to mix states with similar objective value and create the constructive/destructive interference necessary for QAOA performance.
By contrast, the Grover mixer mixes all states equally.
This would be a useful feature for a completely random optimization problem, that is, one where similar states have no correlation in objective value.
In such a case, where there is no structure to the problem, Grover-Th represents an optimal approach as it is effectively just searching via Grover's algorithm, which is known to be optimal for unstructured search.
However, in the case of highly structured optimization problems such as $k$-SAT, the fine-tuned mixing capabilities of the transverse-field mixer, combined with the additional information captured by the objective phase separator (as opposed to the the more brute-force thresholded version) gives the best performance. Our results in effect show how QAOA exploits problem structure to move beyond Grover's unstructured search performance.

Future work~\cite{golden2023} is needed to precisely quantify the different effects and capabilities of these mixers.
Furthermore, it is intriguing to consider mixers custom designed for the specific energy spectra of different optimization problems, e.g. MaxCut vs. Max 2-SAT vs. Max 3-SAT.
The results of this study suggest that brute-force, or maximally mixing QAOA is unlikely to lead to optimal performance for large problem instances.
Instead, mixers and phase separators that are highly tailored to specific problem classes may unlock the fullest of QAOA's potential.

\bibliographystyle{quantum}
\bibliography{references}

\end{document}